# Identity Management issues in Cloud Computing

Smita Saini[1], Deep Mann[2]
*1(Dept. of Computer Science Engineering, Lovely Professional University, India)*
*2(Dept. of Computer Science Engineering, Lovely Professional University, India)*

***Abstract***— Cloud computing is providing a low cost on demand services to the users, omnipresent network, large storage capacity due to these features of cloud computing web applications are moving towards the cloud and due to this migration of the web application, cloud computing platform is raised many issues like privacy, security etc. Privacy issue are major concern for the cloud computing. Privacy is to preserve the sensitive information of the cloud consumer and the major issues to the privacy are unauthorized secondary usage, lack of user control, unclear responsibility. For dealing with these privacy issues Identity management method are used. This paper discusses the privacy issue and different kind of identity management technique that are used for preserving the privacy.

## I. INTRODUCTION

"In 2008 M. Klems said that you can scale your infrastructure on demand within minutes or even seconds, instead of days or weeks, thereby avoiding under-utilization (idle servers) and over-utilization (blue screen) of in-house resources" [1]. Cloud computing model is categorized into deployment model and service model. Deployment model: Deployment model is pointing out the location and management of the cloud infrastructure. The major task of the deployment model is deciding which type of the model is deploying for the service. Deployment model is categorized into the four clouds: Public Cloud, Private Cloud, Hybrid Cloud, community model. Service model is divided into three layer bottom layer, middle layer and top layer these layers are- SaaS (service as a service), PaaS (platform as a service), IaaS (infrastructure as a service). IaaS is a bottom layer of the service oriented architecture (SOA). IaaS is the bottom layer of the service oriented architecture. IaaS layer provide the infrastructure for the cloud consumer. It provide the resource on demand bases, storage space , network and computer resource example- Amazon provides the IaaS, middle layer is PaaS (platform as a service) in this consumer can directly deploy their application without expanding the cost on purchase of the software. Google Apps Engine application is the example of the PaaS provider. The top layer is SaaS (Software as a Service), cloud consumer can directly install application on cloud user are not worry for the infrastructure, Example of SaaS is Salesforce.com [2].

Cloud service provider provides services to the consumer and for the authentication consumer provides the sensitive information to service provider and sometime this information is obtained by the unauthorized user or it used by the service provider for the junk advertisement. Such as Identity management method (IDM) is used for preserve the sensitive information of the consumer. Identity management method enable the user to keep track of the sensitive information that is provided to the service provider (what personal information reveal to the service provider, control how the information can be used, cancel their subscription to the service, and monitor to verify that a service provider applies required policies). Identity management method is used for manage the different digital identity of the consumer and maintain various username/password that is associated with each other.

## II. PRIVACY ISSUES

Cloud computing provides the shared infrastructure to the consumer and due to this, privacy issues is the major challenge. Privacy is major concern to preserve the sensitive information (credit card number, identity information etc…) of the consumer and if such information can be retrieved by the unauthorized user such following privacy issues can be occur [3].

*1. Unauthorized secondary usage-* It is major challenges in the web application like- social networking site are using a user data for advertisement. Same as Cloud service provider are using consumer personal information for the advertisement or junk advertisement. Some time users do not want to reveal his/her personal information.

*2. Lack of user control-* Cloud provides the storage space to the user; users saved his/her data own cloud and users data can be tampered by the service provider, Which mean users loss the control on the data, theft can be occur or users data can be misused by service provider. Such that privacy is compromised of the data, for preserve the privacy we use the privacy protection mechanism.

*3. Unclear responsibility-* This is also a major problem to the privacy, some time users are not aware that which service provider will responsible for the privacy policies, mechanism and privacy protection. Users must be aware who is modifying or accessing his/her data.

## III. PRIVACY MECHANISM TO DEAL WITH THE PRIVACY ISSUES

Identity management provides the solution for unauthorized secondary usage and lack of user control.

*1. Identity management*

Identity is an entity and an entity is a set of unique characteristics. Entity is used for authentication with service provider for accessing the resources. Digital identity is like an Identity card belongs to a particular person that is use for voting, through this person can authenticate themselves. Managing digital identities for human is not an easy task it raises a many security and privacy issues. For managing the digital identity different kinds of identity management technique are used. Identity management identifying a particular person on the bases of claims value (name, email address, credit card number) and prevent the unauthorized access of the resources. Identity management model is based on following things.
*Service provider*- Service provider gives the authority to access the resource to the authenticated user.
*Identity provider*- It issues the digital identity for a user like government issues a pan card.
*Entity* - For whom user are claiming.
*Relying party*- It is used to verify the claims value for whom particular user are claim. Service provider send a request to relying party for verify the claims.

There are several solutions for IDM like isolated IDM model, centric identity model, decentralized identity model, federated IDM, in isolated IDM model service are owned but it is managed by the different service providers and each provider provides service-specific identifier and does identity management by themselves. These are centralized and decentralized approach centralized approach is more significant method. In user centric identity management, each user is assigned several attributes of the identity management. These attributes are identifying the authorized user. Centralized is somewhat similar to the federated IDM. Federated also manage the attributes and credential and authentication and authorizes for entities in its domain. The most widely use solution for the Federate are SAML and Liberty Alliance Standard.

*2. Different Identity Management Methods*
These methods are PRIME (Privacy and identity management for Europe), OpenID, Microsoft Cardspace.

*2.1 PRIME (privacy and identity management for Europe)*
PRIME is using a consol that is handling the service requestor's data. Console requires installation and configuration by using the console service requestor can manage his/her personal information and manage where and what data should be disclosed on the service request.
PRIME provides the authentication by using the anonymous credentials. Users request for the claims to the service provider. A major limitation of PRIME is that it requires user agents and SPs to implement the PRIME middleware [4].

*2.2 OpenID*
For access many web application OpenID user need only one username and one password. Every user has his/her OpenID for authenticate them and use the token to authenticate to web application.
OpenID does not need to provide sensitive information such as user name, credit card information, email address etc… OpenID is a decentralized approach so service provider is not need to register [4]. An end user can freely choose the OpenID provider and if they switch OpenID provider there user no need to provide the credential information. OpenID is highly at risk of phishing attacks.

*2.3 Microsoft Cardspace*
For identifying and verifying Window Cardspace is used the claims value, Claims are like gender, name etc… These claims are used as digital identities to the internet users. It is plug-in with Internet explorer 7 browser and windows XP [5]. In Cardspace digital identity transmitted in form of security token and security token consists of the user private information like user name, home address, SSN number and credit card numbers. These claims are used to prove that it is belongs to the particular users who are claiming.

In identity management system three parties are involves first one is Identity provider (IDP); it is basically issues the digital identity ex- credit card provider issues the digital identities for enabling the payment. Second is Relying party, it provides a service to user for provide the service to user it requires the digital identities. Third is User or service requester, it request for the service from the relying parties and for whom claims are made. In Windows Cardspace, security token is using in the form of SAML [4]. Major limitation of the Window Cardspace is relying on single layer authentication and second is relying on the third party [5].

## IV. REVIEW OF PRIVACY ISSUES

**1.** Xing Wu 2012; author described that centralized IDM is facing the bottleneck problem and federated IDM is based on the single sign on decentralized IDM is also based on the federated IDM but decentralized IDM using the concept of group it make the group on the bases of their relationship if services are having a tight relationship they make one group means they are communicating frequently and these group are called TC (trust context). If any incantation occurs between TCs we create another TC in higher level until we get TCs that meet our criterion. For creating a group it used grouping algorithm [2].

**2.** Bharat Bhargav 2011; authors described about the Windows Card space identity management technique and other identity management technique. Windows card space is based on the federation technique and for security token it uses a SAML it is in the form of XML it provide the more security. Existing Cardspace is having two limitations, trusts on relying party and single layer authentication. For overcome these limitation author used the zero based knowledge technique (ZKP). ZKP integrate with SAML token containing the value of claim. By using the ZKP in security token user satisfied the relying party policy but does not disclose any credential information. But still it depends on the Relying party and token generation and saving is also an issue for windows Cardspace [4].

**3.** Pelin Angin; author described about the entity centric approach. Entity centric approach is based on the active bundle and anonymous identification. An active bundle is kind of container and it consist sensitive data, metadata, and a virtual machine (VM). Sensitive data is a data that is protect from unauthorized access. Meta data described the active bundle and its privacy policies and virtual machine manage and control the program. User can decide what and when data will be share or disclose [6].

**4.** Walled A. Alrodhan; author described the limitation of windows card space these limitations are Single layer authentication to the identity provider and Relying on the third party and the Proposed solution based on the Schnorr's knowledge protocol to the Cardspace. It described some guessing attack that can break the Second layer authentication by guessing some claims for this author recommend the Relying party choose the claims that is hard to guess [7].

## V. Conclusion

This paper provide the overview of the privacy in cloud computing and the various issues of privacy and discuss the solution for 'Privacy Preserving'. Identity management (IDM) method is use for deal with Privacy issues. It gives the overview of different kind of identity management solution and the limitation of these methods. Among presented IDMs Microsoft Cardspace is found to be most popular and adoptive in industries. Cardspace is much flexible and many researchers are keen to improve its efficiency but the major limitation of the identity management solution is relying on the Third party. The future work is to improve Microsoft Cardspace and deploy it to the real time environment.